\documentclass[12pt]{article}
\usepackage{epsfig}
\usepackage{amsmath}
\usepackage{cite}
\usepackage{amssymb}
\newlength{\defbaselineskip}
\newcommand{\setlinespacing}[1]%
           {\setlength{\baselineskip}{#1 \defbaselineskip}}

\begin{document}

\title{Comparative study of fusion barriers using Skyrme interactions and the energy density functional}

\author{O. N. Ghodsi, F. Torabi\thanks{Email: f.torabi@stu.umz.ac.ir}\\
\\
{\small {\em  Department of Physics, Faculty of Science, University of Mazandaran}}\\
{\small {\em P. O. Box 47415-416, Babolsar, Iran}}\\
}
\date{}
\maketitle

\begin{abstract}
\noindent Using different Skyrme interactions, we have carried out a comparative analysis of fusion barriers for a wide range of interacting nuclei in the framework of semiclassical Skyrme energy density formalism. The results of our calculations reveal that SVI, SII, and SIII Skyrme forces are able to reproduce the empirical values of barrier heights with higher accuracy than the other considered forces in this formalism. It is also shown that the calculated nucleus-nucleus potentials derived from such Skyrme interactions are able to explain the fusion cross sections at energies near and above the barrier.

\end{abstract}

\newpage
\setlinespacing{1.5}
\noindent{\bf{I. Introduction}\\}

Fusion of interacting nuclei and associated phenomena have attracted a considerable number of studies to date \cite{1,2,3,4,5,6,7,8}. To analyze such a process between different combinations of target and projectile nuclei, an accurate knowledge of ion-ion interaction potential, especially around the barrier, plays a significant role. The total interaction potential between colliding nuclei is generally related to the long-range Coulomb repulsion of protons as well as the short-range nuclear attraction of nucleons of the interacting nuclei. Because the nuclear part of the total interaction potential is not as well known as the Coulomb part, proposing a precise method or model would be absolutely essential to describe the fusion barriers. Over the last decades, in low-energy nuclear physics, this has become one of the most important challenges in analyzing the fusion process. Due to the investigations, various methods and models based on microscopic, semimicroscopic, and macroscopic approaches \cite{9,10,11,12,13,14,15,16} have been developed to give a reasonable description of the nuclear potential between colliding nuclei.

Among different methods, the Skyrme energy density functional derived from one of the most successful and popular effective interactions \cite{17,18}, together with the extended Thomas-Fermi model (ETF), has been known as an efficient method to study fusion barriers and fusion cross sections. To evaluate interaction potential by this method, target and projectile density profiles and Skyrme force parameters are considered as the main inputs. In view of this, the proper choice of these factors is of particular importance to perform a successful study of fusion barriers in the  energy density formalism. Since 1972 a large number of parametrizations of the Skyrme effective interaction have been introduced. According to each parameter set of Skyrme forces, neutron and proton densities can be self-consistently determined through different methods such as the microscopic Hartree-Fock (HF) and Hartree-Fock-Bogoliubov (HFB) approximations and the semiclassical ETF method \cite{18,19,20}. Moreover, it has been proven that, in the semiclassical ETF approach, the use of the Fermi density whose parameters are obtained \cite{21} by fitting the experimental data, instead of HF and ETF densities, can be totally reasonable \cite{21}.

So far, self-consistent densities obtained from common methods with some parametrizations of Skyrme forces have been applied to describe the fusion barriers of different combinations of projectiles and targets in the semiclassical ETF method of the Skyrme energy density functional \cite{15,22,23}. Density distributions of nuclei, which are calculated by different methods, and parameter sets of Skyrme interactions have influence on calculations of fusion barriers in the energy density formalism. Considering this point, in the present study, we performed a comparative and systematic analysis of fusion barriers using several Skyrme interactions, together with Fermi density distributions including parameters determined by fitting the experimental data \cite{21,24}, in the semiclassical energy density formalism. The obtained results show which Skyrme forces, with these density parameters, are able to more accurately describe the fusion barrier heights. In addition, using the calculated potentials, the theoretical fusion cross sections were computed and compared to the corresponding experimental data.

This paper is organized as follows: Section II gives a brief description of the semiclassical expression of the Skyrme energy density functional. Section III contains the results of the calculations of fusion barrier characteristics and fusion cross sections. Finally, Sec. IV is devoted to a summary of the study.
\\

\noindent{\bf {II. THEORETICAL OUTLINE }}\\

\noindent{\bf\small{1. Skyrme energy density functional in the semiclassical ETF approximation}}

In the energy density functional, the nuclear part of the interaction potential, $V_{N}(R)$, between projectile and target nuclei is defined as a function of $R$, the distance between the center of mass of nuclei, by the following form:

\begin{equation} \label{1}
V_{N}(R)=E_{12}(R)-(E_{1}+E_{2}),
\end{equation}
\begin{equation} \label{2}
E_{12}(R)=\int{\mathcal{E}\left[ \rho_{1p}(\vec{r})+\rho_{2p}(\vec{r}-\vec{R}),\rho_{1n}(\vec{r})+\rho_{2n}(\vec{r}-\vec{R})\right] \mathrm{d^3}r,}
\end{equation}
\begin{equation} \label{3}
E_{1}=\int{\mathcal{E}\left[ \rho_{1p}(\vec{r}),\rho_{1n}(\vec{r})\right] \mathrm{d^3}r,}
\end{equation}
\begin{equation} \label{4}
E_{2}=\int{\mathcal{E}\left[ \rho_{2p}(\vec{r}),\rho_{2n}(\vec{r})\right] \mathrm{d^3}r.}
\end{equation}
\\
In the above equations, the frozen neutron and proton densities of colliding nuclei are shown by $ {\rho_{1n}, \rho_{2n}, \rho_{1p}}$, and ${\rho_{2p}} $, and the Skyrme energy density has the following expression:

\begin{equation}
\begin{aligned}
\mathcal{E}(\vec{r})= & \frac{\hbar^{2}}{2m}\tau+\frac{1}{2}t_{0}\left[\left( 1+\frac{1}{2}x_{0}\right) \rho^{2}-\left(x_{0}+\frac{1}{2}\right) ({\rho_{n}}^2+{\rho_{p}}^2)\right] \\
 &+\frac{1}{12}t_{3}\rho^{\alpha}\left[\left( 1+\frac{1}{2}x_{3}\right) \rho^{2}-\left( x_{3}+\frac{1}{2}\right) ({\rho_{n}}^2+{\rho_{p}}^2)\right] \\
 &+\frac{1}{4}\left[t_{1}\left( 1+\frac{1}{2}x_{1}\right) +t_{2}\left(1+\frac{1}{2}x_{2}\right) \right] (\rho\tau)\\
 &-\dfrac{1}{4}\left[t_{1}\left(x_{1}+\dfrac{1}{2}\right) -t_{2}\left( x_{2}+\frac{1}{2}\right) \right] (\rho_{n}\tau_{n}+\rho_{p}\tau_{p}) \\
 &+\frac{1}{16}\left[3t_{1}\left(1+\frac{1}{2}x_{1}\right) -t_{2}\left(1+\frac{1}{2}x_{2}\right) \right](\vec{\nabla}\rho)^{2}\\
 &-\frac{1}{16}\left[3t_{1}\left(x_{1}+\frac{1}{2}\right) +t_{2}\left(x_{2}+\frac{1}{2}\right) \right] ((\vec{\nabla}\rho_{n})^{2}+(\vec{\nabla}\rho_{p})^{2})\\
 &+\frac{1}{2}W_{0}\left[\vec{J}.\vec{\nabla}\rho+\vec{J_{n}}.\vec{\nabla}\rho_{n}+\vec{J_{p}}.\vec{\nabla}\rho_{p}\right] .
 \end{aligned}
\end{equation}
\

In Eq. (5), $m$ is the nucleon mass. $\rho=\rho_{n}+\rho_{p}$, $\tau=\tau_{n}+\tau_{p}$, and $\vec{J}=\vec{J_{n}}+\vec{J_{p}}$ are defined as nuclear, kinetic energy, and spin-orbit densities, respectively, and $ x_{i} $, $ t_{i} (i=0,1,2,3)$, $\alpha$, and $ W_{0}$ are Skyrme interaction parameters. So far, many parametrizations of Skyrme forces have been introduced in different studies. In this work, some of them, which can successfully describe properties of nuclei, have been selected for the study of fusion barriers in the described formalism.

Considering the semiclassical correction of the second order in the ETF model, the kinetic energy density is defined as a function of local density and its derivations $(q=n$ or $ p)$,

\begin{equation}\label{6}
\begin{aligned}
{\tau}^{(ETF)}_{q}(\vec{r}){}=&\frac{3}{5}(3\pi^{2})^{\frac{2}{3}}{\rho_{q}}^{\frac{5}{3}}+\dfrac{1}{36}\frac{({\vec{\nabla}{\rho_{q}})}^{2}}{\rho_{q}}+\dfrac{1}{3}\Delta\rho_{q}+\frac{1}{6}\frac{\vec{\nabla}\rho_{q}.\vec{\nabla}f_{q}}{f_{q}}+\dfrac{1}{6}\rho_{q}\frac{\Delta{f_{q}}}{f_{q}}-\frac{1}{12}\rho_{q}\left( \frac{\vec{\nabla}f_{q}}{f_{q}}\right) ^{2}\\
&+\frac{1}{2}\rho_{q}\left( \frac{2m}{\hbar^{2}}\right)^{2}\left( \dfrac{W_{0}}{2}\frac{\vec{\nabla}{(\rho+\rho_{q})}}{f_{q}}\right) ^{2},
\end{aligned}
\end{equation}
with the effective mass form factor $f_{q}(\vec{r})$,
\\
\begin{equation} \label{7}
f_{q}(\vec{r})=1+\dfrac{2m}{\hbar^{2}}\frac{1}{4}\left[t_{1}\left(1+\frac{x_{1}}{2}\right) +t_{2}\left( 1+\frac{x_{2}}{2}\right) \right] \rho(\vec{r})-\dfrac{2m}{\hbar^{2}}\frac{1}{4}\left[t_{1}\left( x_{1}+\frac{1}{2}\right) -t_{2}\left( x_{2}+\frac{1}{2}\right) \right]\rho_{q}(\vec{r}).
\end{equation}

As spin $ (\vec{J} ) $ is a purely quantum mechanical feature that has no classical counterpart, the semiclassical expansion of $ \vec{J} $ starting at the second order of $ \hbar $ has the following form:

\begin{equation} \label{8}
{\vec{J}}_{q}(\vec{r})=-\dfrac{2m}{\hbar^{2}}\dfrac{1}{2}W_{0}\frac{1}{f_{q}}\rho_{q}\vec{\nabla}(\rho+\rho_{q}).
\end{equation}

It should be noted that to describe the total ion-ion potential, the Coulomb interaction between colliding nuclei is added to Eq. (1) by using the following formula:
\\
\begin{equation} \label{9}
V_{C}(R)=\int\dfrac{{\rho}^{(1)}_{ch}(\vec{r_{1}}){\rho}^{(2)}_{ch}(\vec{r_{2}})}{\vert\vec{r_{1}}-\vec{r_{2}}\vert}\mathrm{d^3}{r_1}\mathrm{d^3}{r_2}.
\end{equation}
\\
In our calculations we assumed $\rho^{(i)}_{ch}\approx e\rho^{(i)}_{p}$.\\
\\

\noindent{\bf\small{2. Density distributions of nuclei}}

Once the proton and neutron densities of nuclei are determined, the ion-ion interaction potential can be obtained using the above
described method. Density distributions of target and projectile nuclei are predicted through different approximations. Of these
methods, the quantum-mechanical self-consistent HF and HFB methods precisely determine densities using Skyrme effective interactions in
the completely microscopic approach. The ETF approximation is considered another method that describes density distribution in a
semiclassical approach. Moreover, in Ref. \cite{21} a comparison drawn among nuclear density distributions obtained by ETF and HF
calculations and the proposed Fermi density in Ref. \cite{21} showed that such a Fermi density \cite{21}, instead of self-consistent
densities, can be reasonably employed in the semiclassical ETF model. This density distribution is given in the following form:
\\

\begin{equation} \label{10}
\rho_{i}(r)=\rho_{0i}{\left[ 1+exp\frac{r-R_{0i}}{a_{i}}\right] }^{-1},
\end{equation}

\noindent{\text{where $\rho_{0i}$  is the central density,}
\begin{equation} \label{11}
\rho_{0i}=\frac{3A_{i}}{4\pi{R_{0i}^3}}{\left[ 1+\frac{\pi^{2}{a_{i}}^2}{{R_{0i}}^2}\right] }^{-1},
\end{equation}
and $ R_{0i} $ and $ a_{i} $ are half-density radii and the surface thickness parameters, which were determined based on fitting the experimental values to the polynomials in the nuclear mass region A = 4-209 as \cite{21}

\begin{equation} \label{12}
R_{0i}=0.90106+0.10957A_{i}-0.0013{A_{i}}^2 +7.71458\times{10^{-6}}{A_{i}}^3-1.62164\times10^{-8}{A_{i}}^4,                                                                                                          \end{equation}
\begin{equation} \label{13}
a_{i}=0.34175+0.01234A_{i}-2.1864\times 10^{-4}{A_{i}}^2+1.46388\times 10^{-6}{A_{i}}^3-3.24263 \times 10^{-9} {A_{i}}^4.
 \end{equation}
\

\noindent{In another study these parameters, $R_{0i}$ and $a_{i}$, were extended up  to A=238 as \cite{24}}
\

\begin{equation} \label{14}
R_{0i}=0.9543+0.0994A_{i}-9.8851\times 10^{-4}{A_{i}}^2+4.8399\times 10^{-6}{A_{i}}^3-8.4366 \times 10^{-9}{A_{i}}^4,                                                                                                           \end{equation}
\begin{equation} \label{15}
a_{i}=0.3719+0.0086A_{i}-1.1898\times 10^{-4} {A_{i}}^2+6.1678\times 10^{-7}{A_{i}}^3-1.0721\times 10^{-9}{A_{i}}^4.                                                                                                          \end{equation}
\

In our calculations, for nuclear density, $ \rho_{i} $, we have used two-parameter density distributions including parameters determined by Eqs. (12) and (13) and Eqs. (14) and (15), which are labeled as case (a) and  case (b), respectively. Substituting each of them, i.e., case (a) and case (b), in Eq. (5), we calculated the nuclear part of the total potential by different parametrizations of Skyrme forces. The selected Skyrme interactions are as follows: SI, SII \cite{18}; SIII, SIV, SV, SVI \cite{25}; SGI, SGII \cite{26}; SkM* \cite{27}; SkT1, SkT5, SKT9 \cite{28}; SkP \cite{29}; SLy4, SLy7 \cite{30}; BSK2 \cite{31}; and BSK14 \cite{32}.
\\

\
\noindent{\bf{III. CALCULATIONS AND RESULTS }\\}

Adding the Coulomb potential, which was calculated by Eq. (9), to a nuclear part, we determined the total interaction potential $ V_{T}(R) $. Considering the total potential, the values of fusion barrier characteristics, i.e., barrier height, $ V_{B} $, and position, $R_{B}$, can be extracted based on the following conditions:

\begin{equation} \label{16}
\left(\frac{\mathrm{d}V_{T}(r)}{\mathrm{d}r}\right)_{r=R_{B}}=0  \quad  \textrm{and}
\quad  \left(\frac{\mathrm{d^2}V_{T}(r)}{\mathrm{d}r^2}\right)_{r=R_{B}}\leq 0.
\end{equation}
\\
In the present study, using the selected Skyrme forces as well as the density distributions of case (a) and case (b), in the energy density formalism, we performed different systematic studies of the barrier heights for a large number of target and projectile combinations including light, medium-light, and medium-heavy systems. Our investigations cover symmetric as well as asymmetric nuclei, which all have been assumed to be spherical in nature. The accuracy of the obtained results of our calculations has been checked by making a comparison between the theoretical barrier heights and the corresponding empirical values \cite{33,34,35,36,37,38,39,40,41,42,43,44,45,46,47,48,49,50,51,52,53,54,55,56,57,58,59,60,61,62,63,64,65,66,67} via the following formula:

\begin{equation} \label{17}
\chi(V_{B})(\%)={\sqrt{\dfrac{1}{N}\sum\limits_{i=1}^N \left( \dfrac{(V^{Calc.}_{B})_{i}-(V^{Emp.}_{B})_{i}}{(V^{Calc.}_{B})_{i}+(V^{Emp.}_{B})_{i}}\right) ^2}}{\ \times100},
\end{equation}
\

\noindent{where $\chi$ explains the relative error of the calculations of barrier heights and $N$ defines the number of studied systems.}

According to the employed nuclear densities and Skyrme interactions in the semiclassical formalism, the theoretical percentage of relative errors of the barrier height calculations are displayed in Fig. 1. In this figure, to show the effect of varying the parameter set of the Skyrme interaction, the calculated percentage of relative errors is plotted as a function of each employed Skyrme interaction. To compare our results with those obtained from the other theoretical model, we also evaluated the barrier heights of the considered interacting nuclei using the Prox. 2010 potential \cite{68}, which, according to the discussion in Ref. \cite{68}, is able to accurately reproduce the fusion barrier characteristics. The obtained theoretical relative error for the barrier height calculations based on this proximity potential is illustrated with the dashed line in Fig. 1.

From the comparison shown in this figure, it is obvious that the employed nuclear density and the parameter set of the Skyrme interaction influence the performance of the semi-classical model. It is also seen that SVI, SII, and SIII Skyrme interactions, together with both cases (a) and (b), can be well applied to predict the barrier heights of the colliding systems in the energy density formalism. To be more precise, calculations based on the SVI Skyrme interaction resulting in $ \chi=1.5079\% $ as well as SII and SIII Skyrme forces with $ \chi=1.5144\% $ and $ \chi=1.5147\% $, respectively, provide more accurate values of barrier heights than the other calculations, i.e., calculations made using the other Skyrme interactions and the considered proximity potential here. Therefore, To achieve a better comparison between the theoretical values of barrier characteristics determined by the three selected forces and the empirical estimates, in Figs. 2 and 3, we show the percentage difference of barrier heights , $ \Delta{V_{B}}(\%) $, and positions, $ \Delta{R_{B}}(\%) $, which are defined as

\begin{equation} \label{18}
\Delta{V_{B}}(\%)=\dfrac{V^{Calc.}_{B}-V^{Emp.}_{B}}{V^{Emp.}_{B}}\times100,
\end{equation}

\begin{equation} \label{18}
\Delta{R_{B}}(\%)=\dfrac{R^{Calc.}_{B}-R^{Emp.}_{B}}{R^{Emp.}_{B}}\times100.
\end{equation}
\

\noindent{As Fig. 2 indicates, using these forces in the semiclassical method, one can predict the values of barrier heights with an accuracy of about $\pm5\%$. However, the results in Fig. 3 show that the barrier positions can be reproduced with an accuracy of about $\pm10\%$. Such a deviation might be related to a great uncertainty in the measurement of barrier positions.}

It is expected that theories which predict the barrier characteristics with reasonable accuracy provide a fair description of the fusion cross-section data at energies in the vicinity of the barrier. Therefore, we employed the potentials obtained from the above-selected forces to analyze the fusion cross sections by the Wong formula \cite{69}, which has been widely applied to explain the fusion cross sections at energies near and above the barrier. According to this formula \cite{69}, the fusion cross section is given by the following expression:

\begin{equation} \label{20}
\sigma_{\textrm{fus}}(\textrm{mb})=\frac{10{R_{B}}^{2}\hbar\omega_{0}}{2E_{c.m.}}\ln{\left\lbrace{1+\exp\left[{\dfrac{2\pi}{\hbar\omega_{0}}(E_{c.m.}-V_{B})}\right]}\right\rbrace},
\end{equation}
\

\noindent{where $ \hbar\omega_{l} $ is the curvature of the inverted parabola, $ E_{c.m.}$ is the center-of-mass energy, and $ V_{B} $ is the barrier height.}

The calculated results of the fusion cross sections for some of the colliding systems, \textsuperscript{16}O+\textsuperscript{58}Ni,  \textsuperscript{36}S+\textsuperscript{48}Ca, \textsuperscript{40}Ca+\textsuperscript{40}Ca, and \textsuperscript{19}F+\textsuperscript{197}Au, are compared with the corresponding experimental data \cite{53,70,71,72} in Fig. 4. In this figure, the solid, dashed, and dash-dotted curves are plotted based on the fusion cross sections calculated from the potentials obtained by the SVI, SII, and SIII Skyrme forces, respectively. As it is clearly seen, this comparison demonstrates a good agreement between the theoretical and measured fusion cross sections, at energies near and above the barrier.
 \\

\noindent{\bf{IV. SUMMARY}\\}
\

In this paper, the aim was to analyze the predictions of fusion barrier heights obtained by different Skyrme force parametrizations in the semiclassical expression of the energy density formalism. To this end, employing different Skyrme interactions, together with the nuclear matter density distributions including parameters determined based on fitting the experimental values, we studied the barrier heights of a series of fusion reactions in the energy density functional method. Comparing the theoretical relative errors of the barrier height calculations in Fig. 1, one can conclude that the barrier heights evaluated by using the SVI, SII, and SIII Skyrme interactions, $ \chi\sim1.5\% $, are closer to the empirical estimates. From Fig. 4, it is also evident that the experimental cross sections, at near and above barrier energies, can be reproduced by the potentials obtained from these three Skyrme forces.\\
\\
\\


\newpage
\noindent{\bf {FIGURE CAPTIONS}}\\
\\
"(Color online)" Fig. 1. The percentage relative errors of the calculations of barrier heights using densities of case (a) and case (b) as well as different Skyrme interactions in the semiclassical expression of the Skyrme energy density functional. The theoretical relative error of the calculations of barrier heights in the proximity approach is illustrated with the dashed line as well.
\\
\\
Fig. 2. The percentage difference of the barrier heights $ \Delta{V_B} (\%)$ as a function of $Z_{1}Z_{2}/({A_{1}}^{1/3}+{A_{2}}^{1/3})$ using (a) SVI, (b) SII, and (c) SIII Skyrme interactions in the energy density formalism.\\
\\
\\
Fig. 3. The percentage difference of the barrier positions $ \Delta{R_B} (\%)$ as a function of $Z_{1}Z_{2}/({A_{1}}^{1/3}+{A_{2}}^{1/3})$ using (a) SVI, (b) SII, and (c) SIII Skyrme interactions in the energy density formalism.\\
\\
\\
"(Color online)" Fig. 4. The fusion cross sections as a function of the center-of-mass energy for different colliding systems: (a) \textsuperscript{16}O+\textsuperscript{58}Ni, (b) \textsuperscript{36}S+\textsuperscript{48}Ca, (c) \textsuperscript{40}Ca+\textsuperscript{40}Ca, and (d) \textsuperscript{19}F+\textsuperscript{197}Au. Theoretical values were calculated using potentials obtained from SVI, SII, and SIII Skyrme forces, and the experimental data were taken from Refs.\cite{53,70,71,72}.
\\
\newpage

\begin{figure}
\begin{center}
\includegraphics{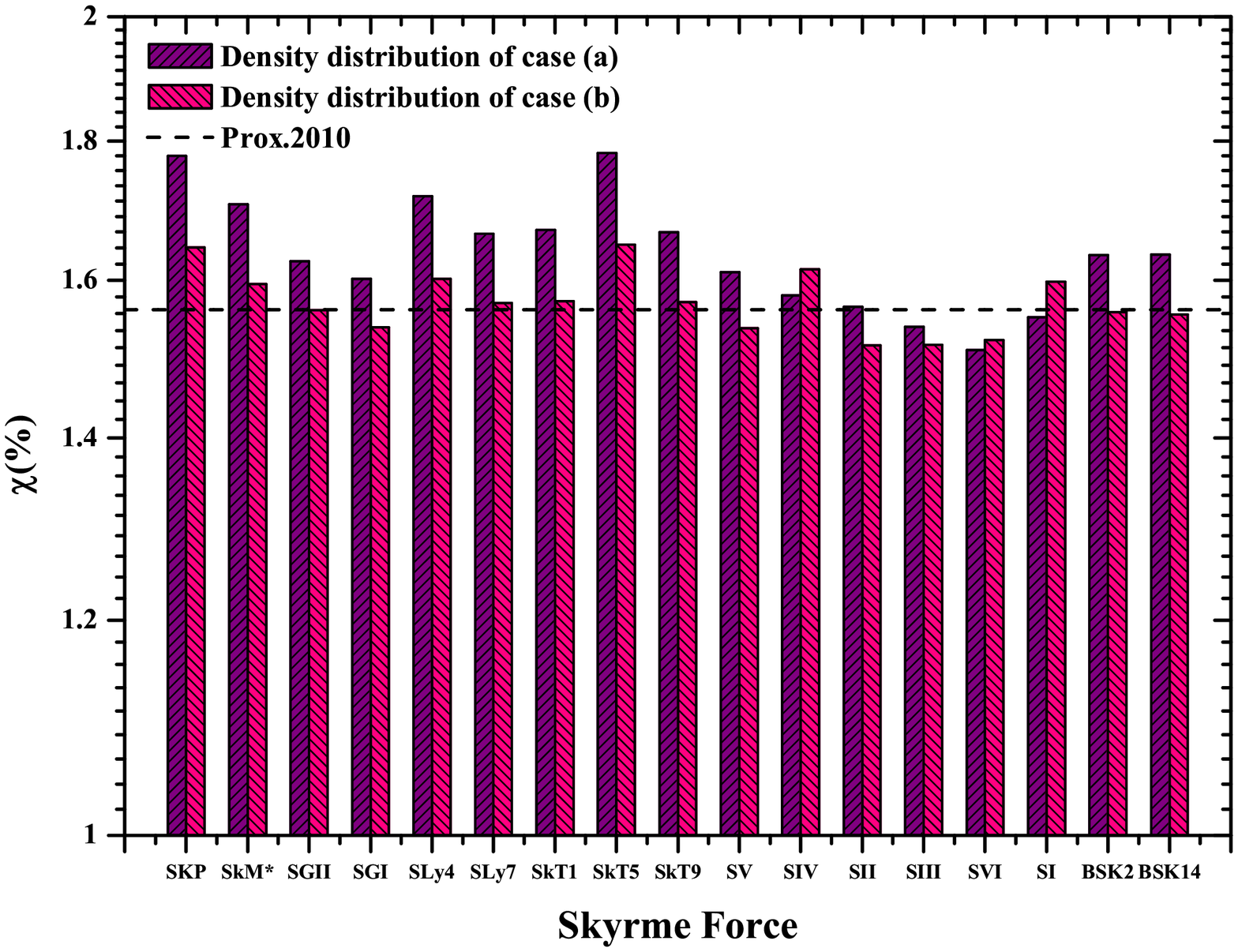}
\end{center}
\vspace{15cm} \caption{}
\end{figure}

\begin{figure}[htp]
  \centering
  \begin{tabular}{cc}
    \includegraphics[width=120mm]{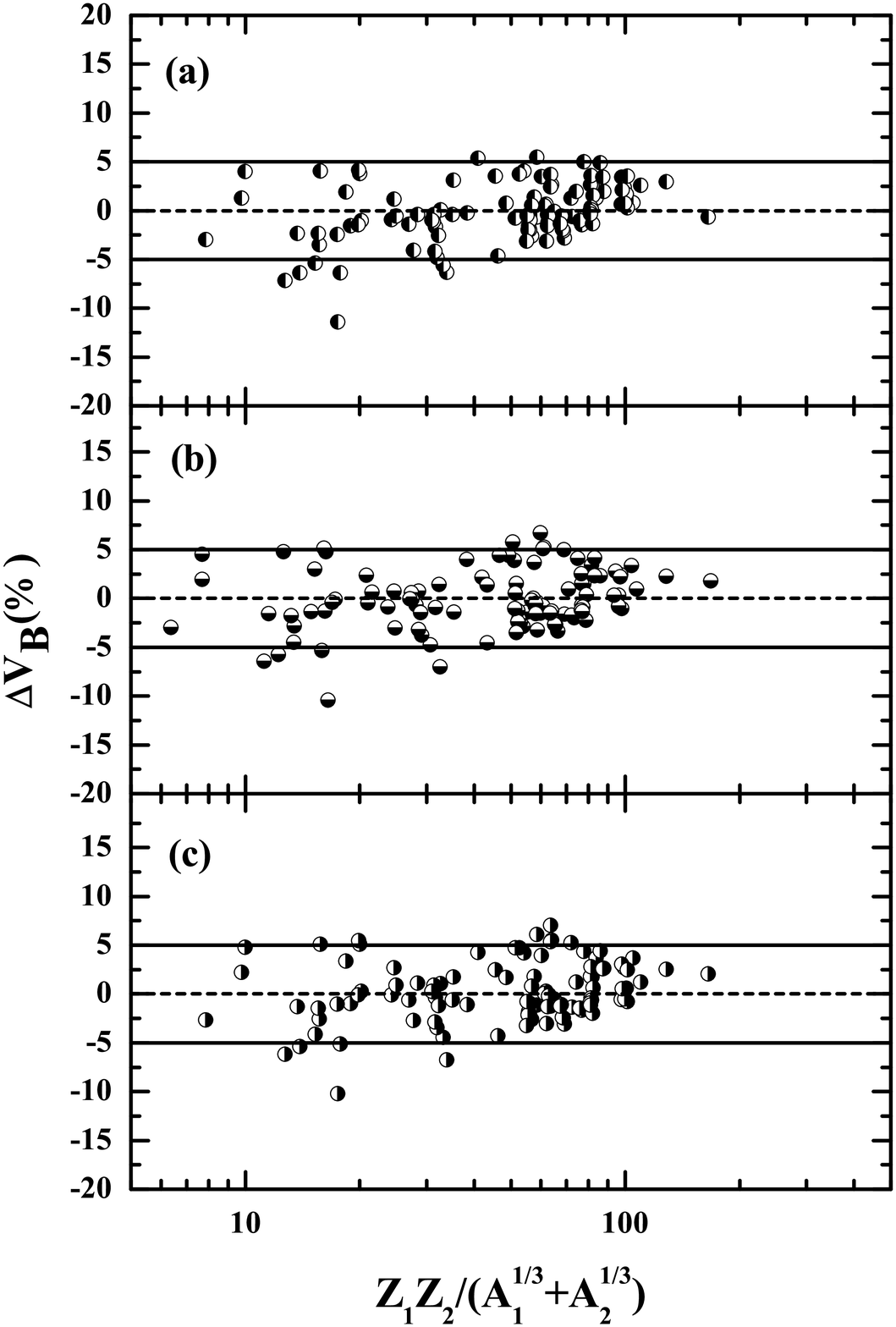}\\
      \end{tabular}
   \caption{}
  \end{figure}

  \begin{figure}[htp]
  \centering
  \begin{tabular}{cc}
    \includegraphics[width=120mm]{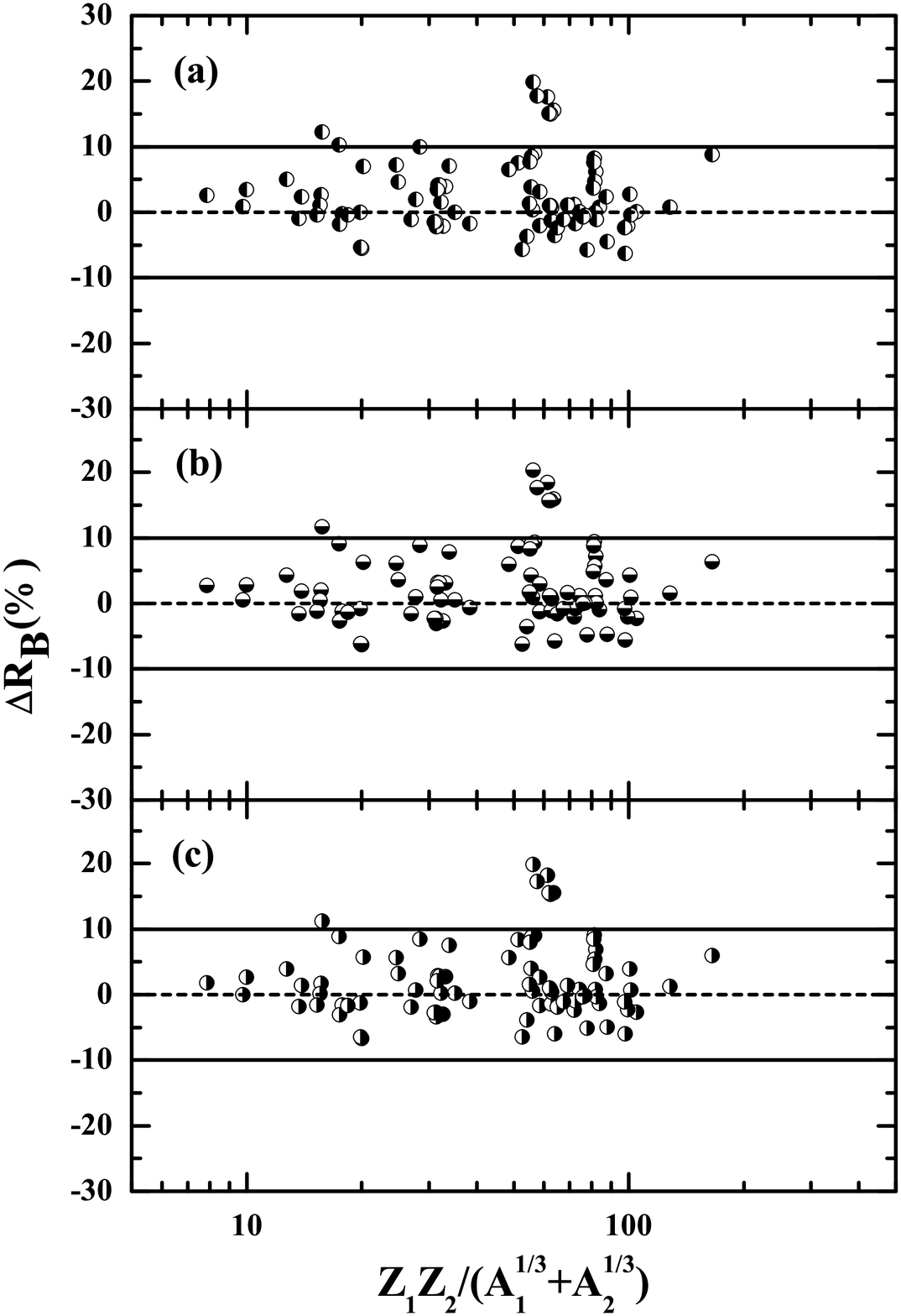}\\
      \end{tabular}
   \caption{}
  \end{figure}

\begin{figure}[htp]
  \centering
  \begin{tabular}{cc}
    \includegraphics[width=86mm]{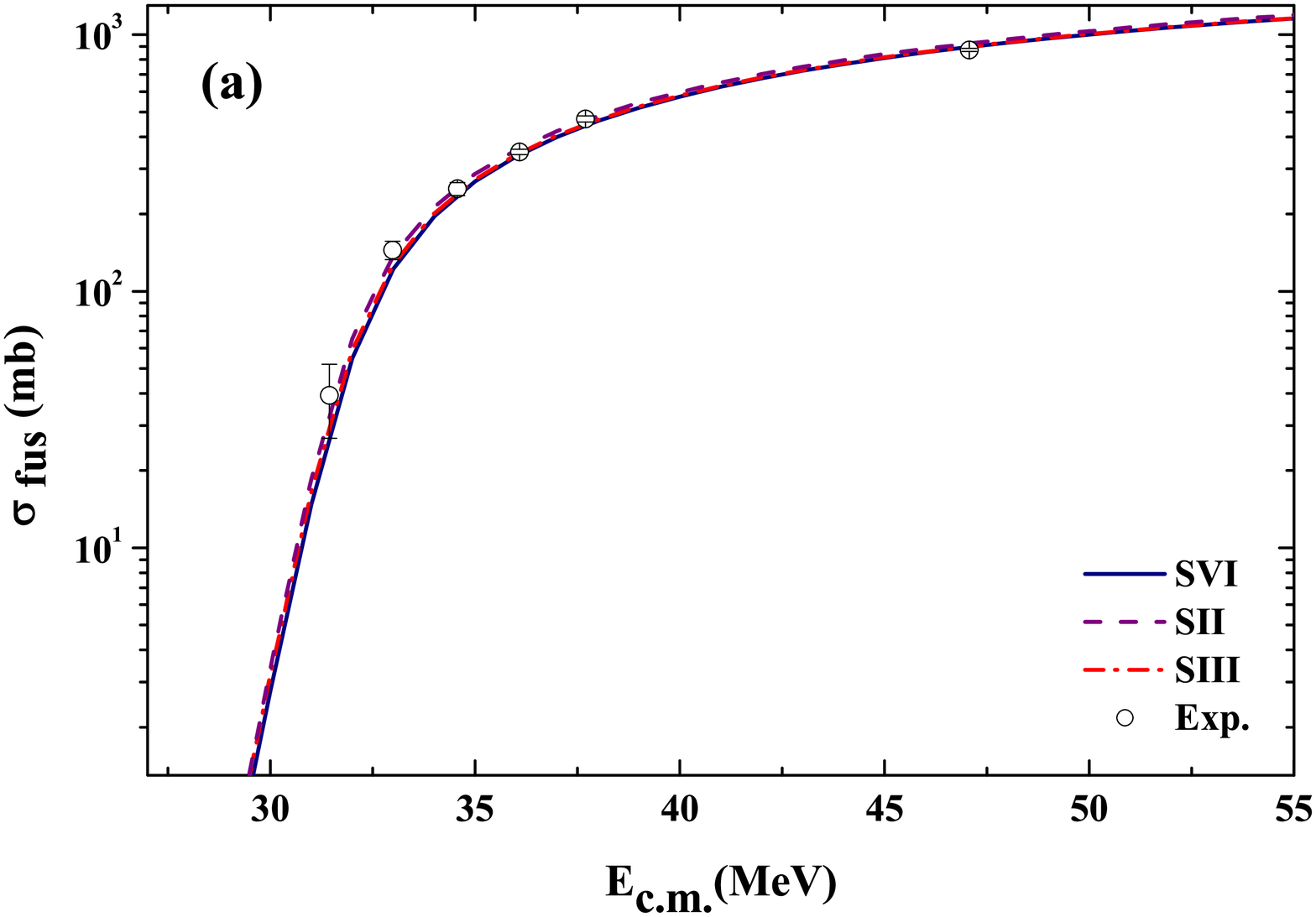}&
    \includegraphics[width=86mm]{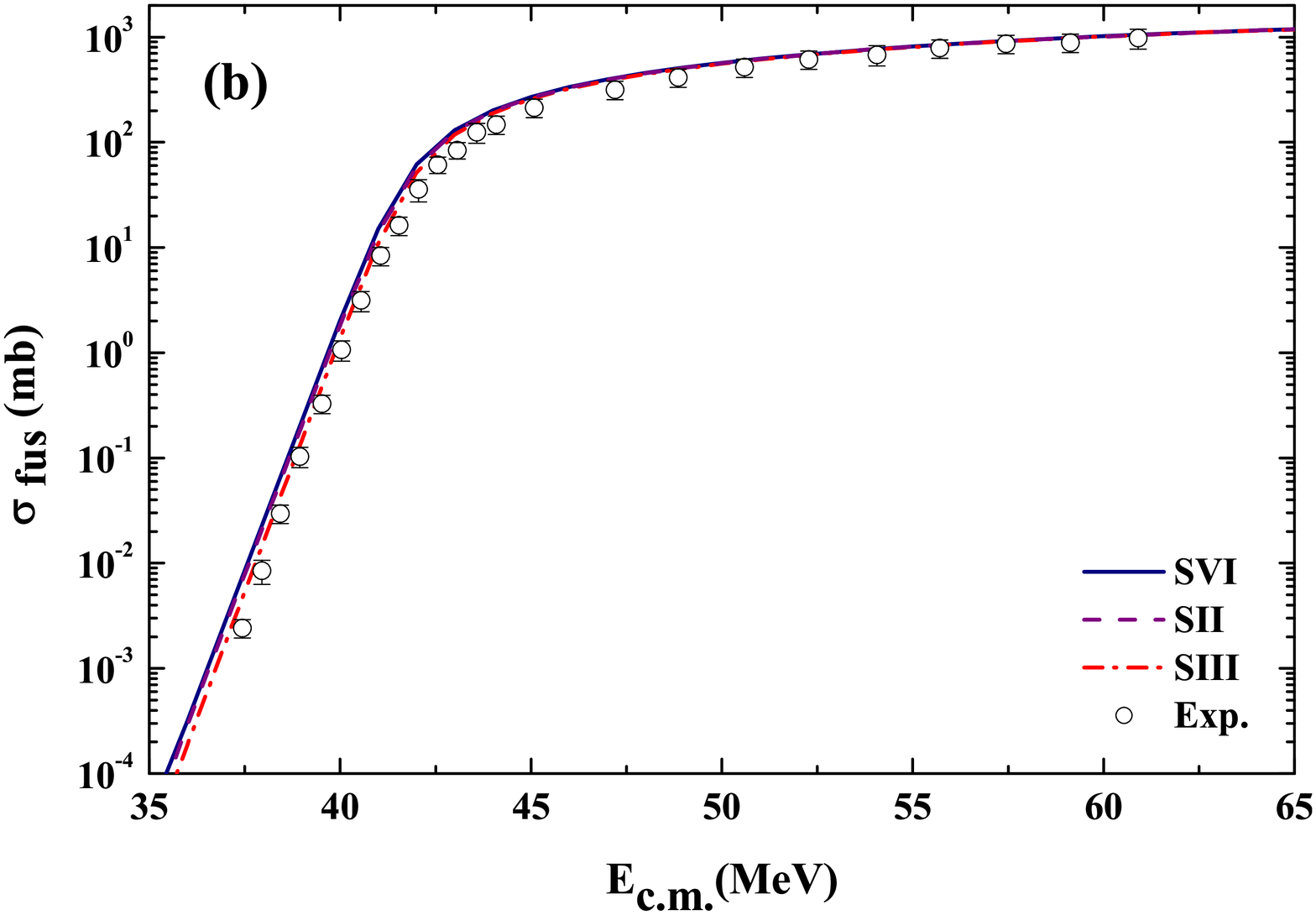}\\
    \includegraphics[width=86mm]{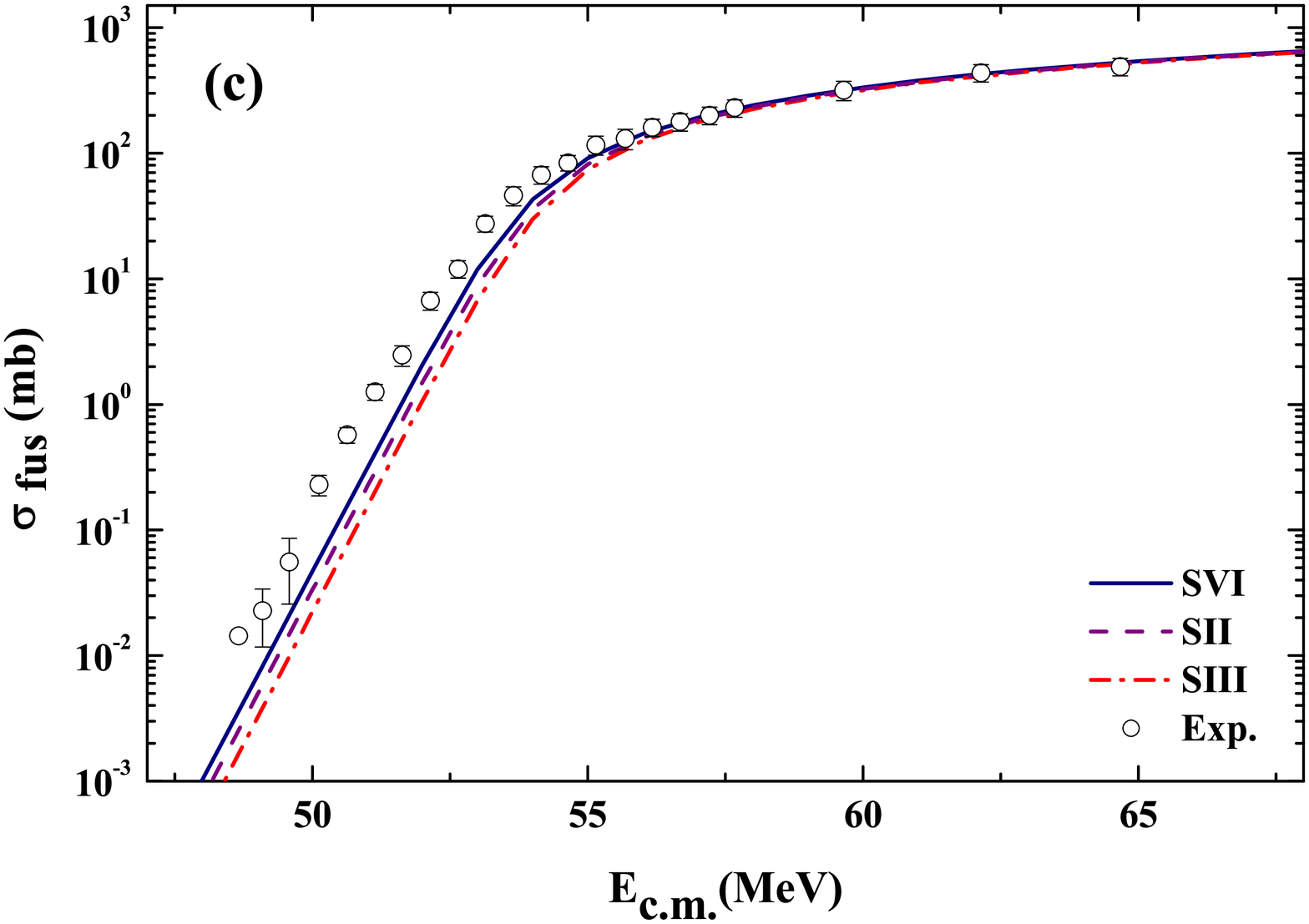}&
    \includegraphics[width=86mm]{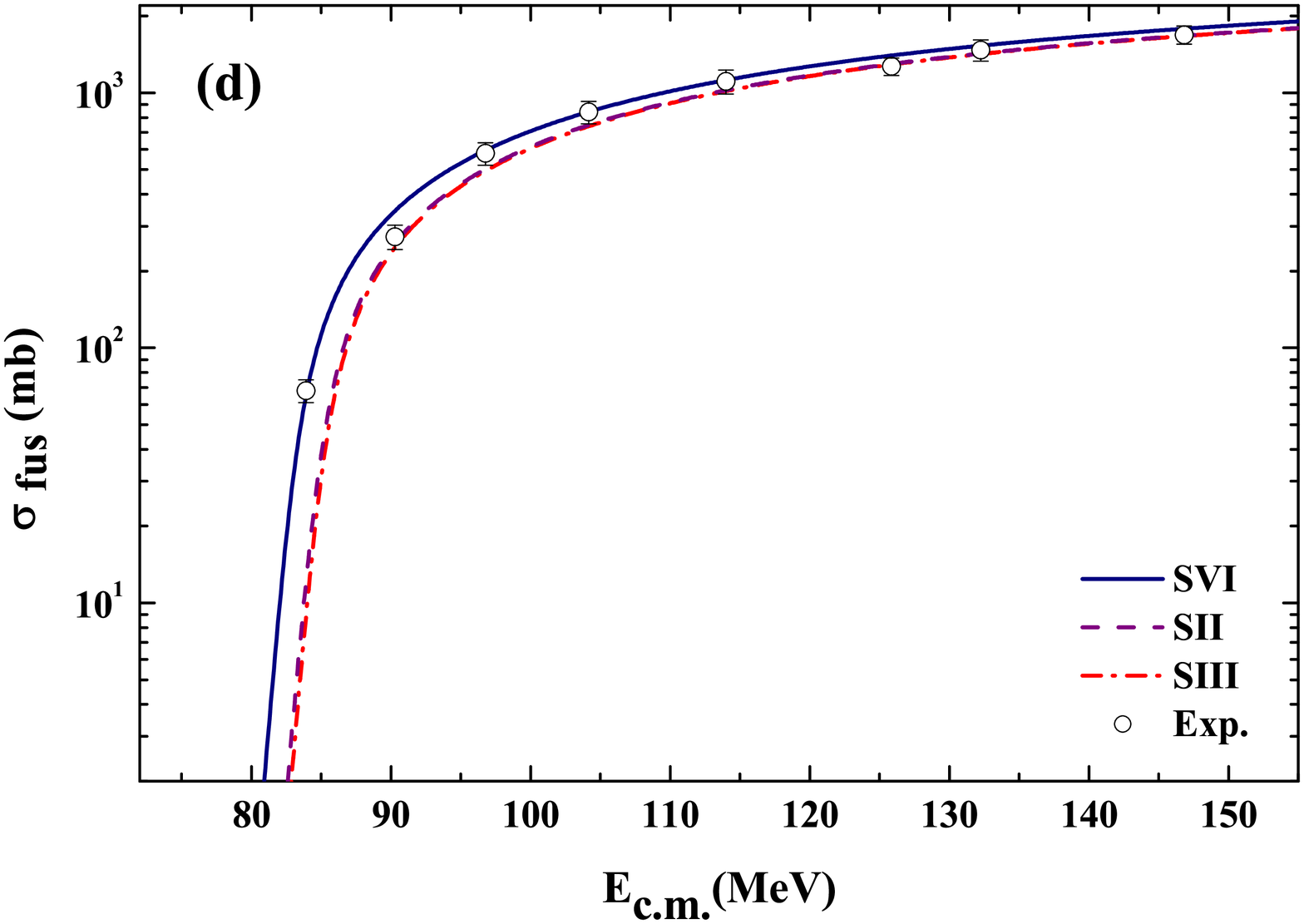}\\
           \end{tabular}
          \caption{}
  \end{figure}

\end{document}